\documentclass[12pt,preprint]{aastex} \usepackage{mathrsfs}

\shorttitle{The Metallicity Distribution
in the Outer Halo of M33} \shortauthors{Brooks et al.}

\begin{document}

\title{The Metallicity Distribution in the Outer Halo of M33}

\author{R. Scott Brooks, Christine
D. Wilson, and William E. Harris}
\affil{Department of Physics and Astronomy, McMaster University,\\
Hamilton, ON L8S 4M1, Canada}


\begin{abstract}
We present the results of deep $I$ and $V$-band photometry of the halo
stars near the southeast minor axis of the Local Group spiral galaxy
M33. An $(I,V-I)$ color-magnitude diagram distinctly reveals the red
giant branch stars at the bright end of the color-magnitude diagram. A
luminosity function in the $I$-band which utilizes a background field
to remove the contaminating effects of Galactic foreground stars and
distant, unresolved galaxies reveals the presence of the tip of the
red giant branch at $(m-M)_{I}=20.70\pm0.10$. Assuming an absolute
magnitude of the tip of the red giant branch of
$M_{I,TRGB}=-4.1\pm0.1$ and a foreground reddening of
$E(V-I)=0.054\pm0.020$, the distance modulus of M33 is determined to
be $(m-M)_\circ=24.72\pm0.14$ ($880\pm60$ kpc). The metallicity
distribution function derived by interpolating between evolutionary
tracks of red giant branch models is dominated by a relatively metal
poor stellar population, with a mean metallicity of
$[m/$H$]=-0.94\pm0.04$ or $[$Fe$/$H$]=-1.24$. We fit a leaky-box
chemical enrichment model to the halo data, which shows that the halo
is well represented by a single-component model with an effective
yield of $y_{eff}=0.0024$. The mean metallicity of a sample of
globular clusters in the M33 halo ([Fe/H]$=-1.3\pm0.1$) matches very
closely the metallicity of the halo, but the cluster sample is too
small to investigate other aspects of the metallicity distribution
function of the clusters.
\end{abstract}




\section{Introduction}\label{intro}

Our current understanding of galaxy formation and evolution has
improved in recent years; however, direct observation of these
processes still lies beyond the ability of the current generation of
telescopes. To better understand these issues, we must rely on
secondary properties that trace galactic formation and evolution. For
the majority of galaxies, globular cluster systems and stellar halos
serve as excellent diagnostic tracers of a galaxy's chemical
evolution. Unfortunately, the majority of galaxies lie at distances too
great to resolve individual stars and so studies of galactic stellar
halos are limited to only the closest galaxies.


The halo of the Milky Way has yielded a wealth of information
regarding the early stages of its chemical evolution. The bi-modal
metallicity distribution of the Milky Way globular cluster system has
been well established, a characteristic shared with many large spirals
and giant ellipticals \citep{har01}. However, being embedded in the
Galactic disk makes studies of individual halo stars difficult due to
significant contamination from disk stars and the requirement of
accurate relative distances to each halo object observed.
Extragalactic globular cluster systems and stellar halos provide an
opportunity to study stellar populations at roughly the same relative
distance, thus eliminating the need for accurate distances to each
halo object. Individual halo stars are easily resolved in M31, the
largest galaxy in the Local Group. The stellar halo and globular
cluster system of M31 are both well studied. \citet{dur01} observed a
large population of red giant branch stars in the halo of M31. Their
resulting metallicity distribution function was bi-modal, dominated by
a moderately high metallicity population ([$m/$H] $\sim{-0.5})$ with a
significant metal-poor population also present. Surprisingly, a
comparison of the stellar halo metallicity distribution function with
that of the globular cluster system reveals they do not match; the
globular cluster system of M31 has a much larger fraction of
metal-poor objects. A similar result is found for the halo of the
giant elliptical galaxy NGC 5128 \citep{har00}. At a distance of 4
Mpc \citep{har00}, individual stars in the halo of NGC 5128 can be
easily resolved with the Hubble Space Telescope.

The Local Group galaxy M33 in Triangulum has been studied relatively
little in comparison to other members of the Local Group. M33 is small
in comparison to M31 and the Milky Way; however, it is typical of
spiral galaxies found throughout the universe. At a distance of $916
\pm 17$ kpc \citep{kim02}, it is relatively straight forward to
perform photometry on the red giant branch stars in the halo of
M33. Although significant for its size, the globular cluster system of
M33 is rather sparse in comparison to the larger members of the Local
Group, M31 and the Milky Way. Early studies of M33's globular cluster
system by \citet{chr82,chr88} revealed a broad spectrum of cluster
ages. More recently, \citet{sar00} derived metallicities for 9
globular clusters in the halo of M33. The mean metallicity of their
sample is $\langle$[Fe/H]$\rangle=-1.27\pm0.11$ dex; however, there is
a significant spread of metallicity in this sample. While the globular
cluster system of M33 has been fairly well studied, very little work
has been done on the stellar population in the halo. The work of
\citet{mou86} represents one of the only studies of this
population. Their analysis of a field 7 kpc from the galaxy center
along the southeast minor axis revealed a very metal poor population,
with a mean metallicity of [{\it m}/H]$=-2.2 \pm 0.8$ ([Fe/H] $\sim$
-1.9 dex using [{\it m}/H] $\sim$ [Fe/H] + 0.3 from
\citet{dur01}). This mean does not compare well with the mean
metallicity of the globular clusters \citep{sar00}. Unfortunately,
photometric incompleteness in their {\it V}-band data limited their
ability to investigate the presence of metal rich stars.

Kinematic studies of the HI extent of M33 have revealed a significant
``integral-shaped'' warp in its disk.  The warp begins at a distance
of 5 kpc from the center of M33 and grows very quickly; at a distance
of 10 kpc from the center, gas is rotating about an axis inclined at
an angle of $40^\circ$ relative to the inner disk \citep{rog76}. Given
that the central disk of M33 has an inclination of $i=55^\circ$
\citep{deu87}, the outer disk is then nearly edge-on (outer disk
inclination of $\sim 95^\circ$). Of particular interest for this work
is the presence of a sharp cut-off of the main HI disc along the
eastern edge of the galaxy \citep{rea78}, since this will also result
in a cutoff in the disk stars.

In this paper, we present new photometry of the stellar halo of
M33. Utilizing the current generation of telescopes, our sample probes
much deeper than previous stellar halo studies of M33, allowing a more
complete investigation of the chemical composition of the halo. In
Section~\ref{sec-obs} we discuss the observational details, subsequent
data reduction, and classification of our data. In
Section~\ref{sec-lf} we derive a luminosity function for the halo
stars and use this to derive the distance to M33. In
Section~\ref{metal} the chemical composition of the halo is
investigated through the derivation of the metallicity distribution
function. The metallicity results are compared to those of the
halo globular clusters and simple chemical evolution models are
employed in an attempt to explain a possible formation scenario for
M33.

\section{Observations}\label{sec-obs}

\begin{figure}
\plotone{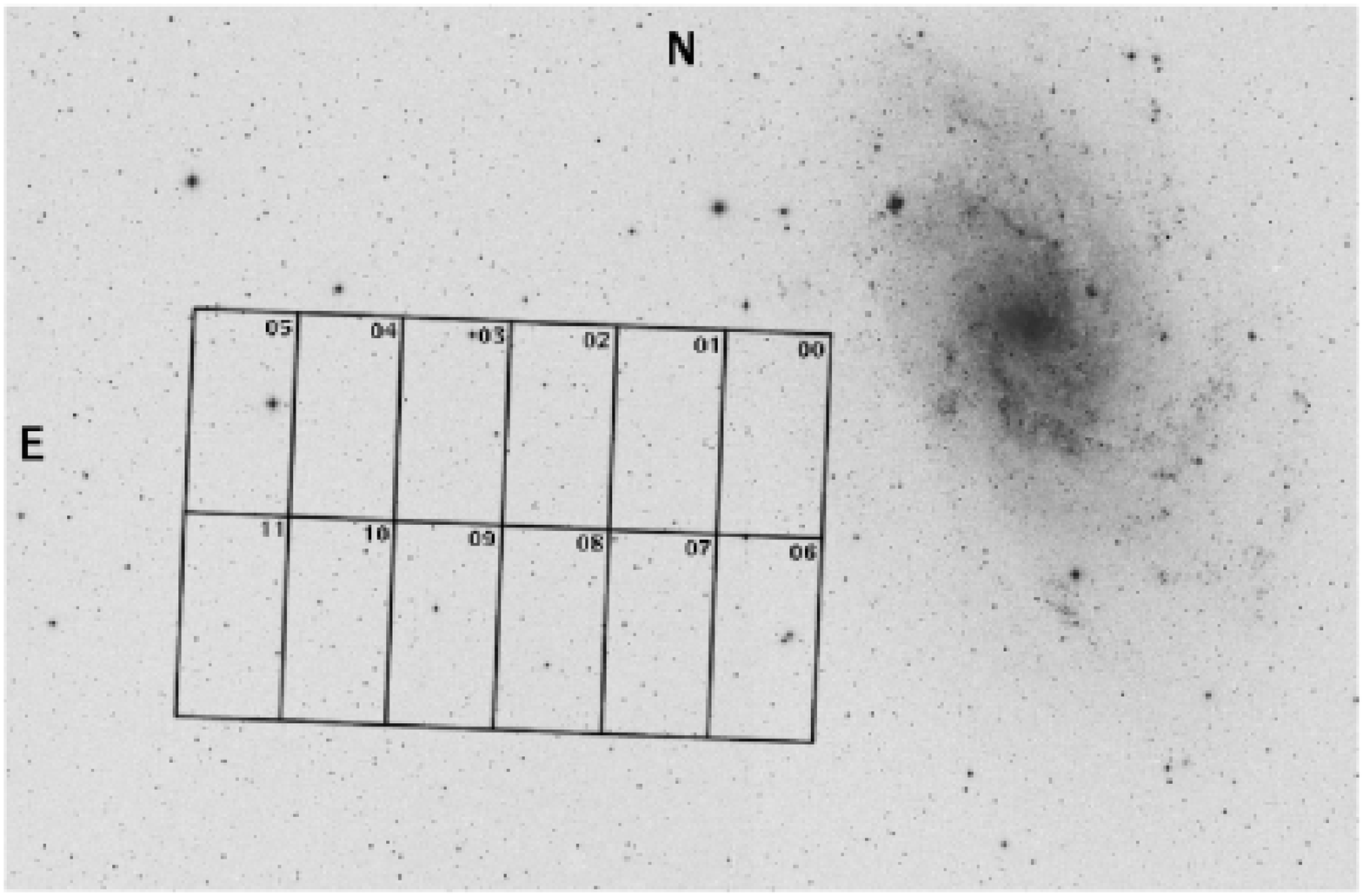}
\caption{Location of the CFH12k mosaic relative to the disk of
M33.\label{m33}}
\end{figure}

$V$ and $I$ band CCD images of a single field in the halo of M33 were
obtained on September 10, 1999 using the CFHT12K camera on the
Canada-France-Hawaii Telescope (CFHT). The CFH12K is a wide-field
mosaic CCD camera, which consists of 12 individual CCD chips, each
2k$\times$4k pixels, arranged in a 2 chip$\times$ 6 chip pattern.  The
pixel size of $15\ \mu{\rm m}$ results in an image scale at the prime
focus of $0.206\ \arcsec \rm{pixel^{-1}}$, which yields a field of
view of $7\times14\ {\rm arcmin^2}$ for an individual chip or
$30\times42\ {\rm arcmin^2}$ for the entire camera.  A total of 6
exposures of M33's halo were obtained in each filter, with individual
exposure times of 900 seconds. Figure~\ref{m33} shows the position on
the sky of the mosaic image relative to the disk of M33\footnote{M33
image obtained from the Digital Sky Survey, made available by the
Space Telescope Science Institute.}.  \notetoeditor{Insert
Figure~\ref{fig-m33} here}

\subsection{Data Preprocessing}

Data preprocessing and calibration were performed with tasks in
IRAF\footnote{Image Reduction and Analysis Facility. IRAF is
distributed by the National Optical Astronomy Observatories, which are
operated by the Association of Universities for Research in Astronomy,
Inc., under cooperative agreement with the National Science
Foundation. {\tt http://iraf.noao.edu/}}. Preprocessing of the program
images (bias and dark subtraction, flat fields, and geometric
correction) involved operations on the entire mosaic with the mosaic
reduction package {\it mscred}. Additional calibration images (twilight
images, darks and bias frames) were obtained from the CADC
archive\footnote{Guest User, Canadian Astronomy Data Center, which is
operated by the Dominion Astrophysical Observatory for the National
Resarch Council of Canada's Herzberg Institute of Astrophysics.}.
Twilight sky images were used to create flat field frames and resulted
in the program images being flattened to the $\sim$1\%-2\% level.

The {\it I}-band images contain significant fringing at the 3\%-4\%
level due to interference within the CCD of atmospheric $OH^-$
emission lines.  Master fringe frames were created for each chip in
the mosaic from the dithered M33 images. The master fringe frame was
then subtracted from each individual chip and exposure. The individual
chips of each exposure were then registered spatially and combined,
with the exposure in each filter with the highest quality seeing being
used as the reference image. 

Large-format cameras, such as the CFH12k, also suffer from significant
geometric distortion across their wide field of view. Rather than
rescale the pixels, the distortion was removed through the use of
correction images made available by the CFHT staff. Dividing the
program images by the correction images rescales the sky level to
produce consistent photometry across the entire mosaic.

\subsection{Calibration}

Photometric calibration was performed with stars in the open cluster
NGC 7790 (Stetson Photometric Standard Fields\footnote{The CADC
Astronomical Standards Page: ~ {\tt
http://cadcwww.dao.nrc.ca/standards/}}).  The majority of the NGC 7790
field was imaged on a single chip of the mosaic (chip 02). The
standards were fit to the usual transformation equation:
\begin{equation} \label{traneqn}
M_n=m_n+a_n+b_nX+c_n(V-I)+\Delta V_{2,n}
\end{equation}
where $M_n$ is the calibrated magnitude, $m_n$ is the aperture
corrected instrumental magnitude (using an aperture radius of 15
pixels), $a_n$ is the zero-point correction, $b_n$ is the extinction
coefficient, $c_n$ is the colour coefficient, $X$ is the effective
airmass of the exposures, and $(V-I)$ is the instrumental colour, all
with respect to chip $n$.
An average value of the extinction coefficient was adopted from
\citet{lan92}. The same color coefficient was used for each chip,
based on the solution from chip 2, and a simple zero-point correction
determined from the relative sky levels between chip 2 and the
remaining chips was applied to the other chips in each filter,
represented by $\Delta V_{2,n}$ in the transformation equation. The
resulting values for the various coefficients are summarized in
Table~\ref{tbl-values}.  \notetoeditor{Insert Table~\ref{tbl-values}
here.}  The transformation equations also require the use of a single
airmass value, which was determined from $X_{eff} =
\frac{1}{6}(X_{init} + 4X_{mid} + X_{end})$ \citep{ste89}.

\subsection{Photometry and Classification}\label{sec-class}

All photometric reductions on the final M33 images were performed with
stand-alone versions of the DAOPHOT II and ALLSTAR packages
\citep{ste87,ste90,ste92}. A first pass with DAOPHOT II and ALLSTAR
with a detection threshold of 6 $\sigma$ above sky was performed,
followed by a second pass, which did not add significantly to the
number of detections. A stellar point spread function (PSF) was
derived for each chip individually with 10 to 20 bright and isolated
stars on each chip. The PSF did not vary significantly across a single
chip and so a constant PSF was used for each chip. Aperture
corrections were defined with 5 to 10 bright, isolated stars on each
chip with an aperture radius of 15 pixels.

\begin{figure}
\plotone{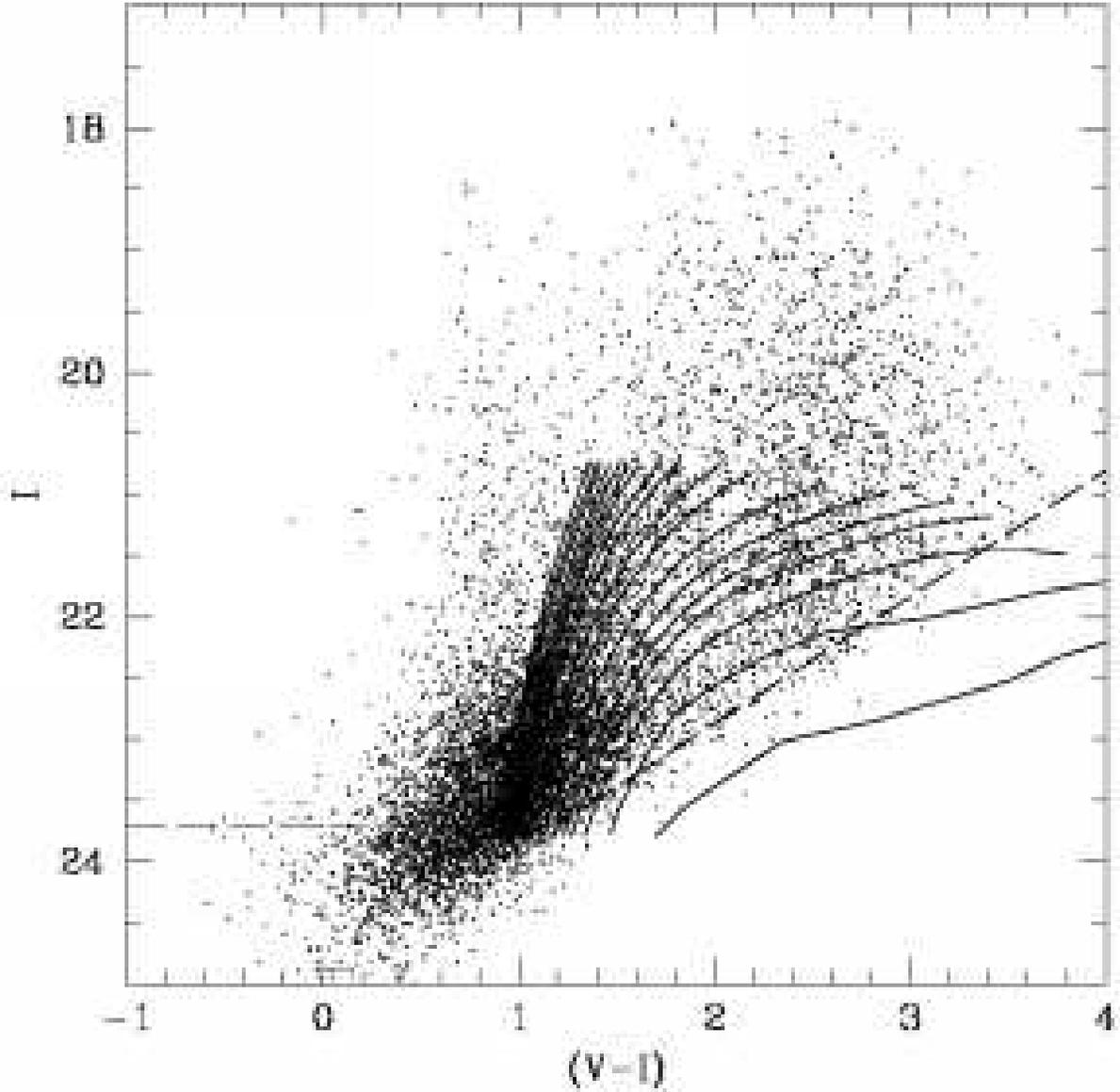}
\caption{$I,(V-I)$ color magnitude diagram of the M33 halo field after
rejection of all nonstellar objects. The dashed line marks the average
50\% completeness level and solid lines show the theoretical models
from \citet{van00}. The two CCD chips closest to the center of M33
suffer from substantial contamination by disk stars and are not
included here.\label{m33cmd}}
\end{figure}

The reduction routines employed by DAOPHOT II discard any obviously
nonstellar objects; however, many slightly resolved background
galaxies can still be present. Removal of these objects was
accomplished with the $\chi$ parameter from DAOPHOT II \citep{ste87}
and the $r_1$ and $r_{-2}$ radial moments \citep{kro80,har81}. The $\chi$
parameter measures the PSF fitting quality, while the $r_1$ and
$r_{-2}$ radial moments examine the extended distribution and the
central concentration of the light, respectively. Galaxies have a more
extended distribution and are less centrally peaked than stars. The
range of $\chi$, $r_1$, and $r_{-2}$ values that defined the stellar
population were derived for each chip in each filter using the results
of artificial star experiments (see Section~\ref{sec-expt}). Only
objects that met these criteria were retained for the remainder of the
analysis. The resulting {\it I,(V-I)} color-magnitude diagram of the
M33 halo field is shown in Figure~\ref{m33cmd}.

\notetoeditor{Insert Figure \ref{m33cmd} here}

\subsection{Artificial Star Experiments}\label{sec-expt}

\begin{figure}
\plotone{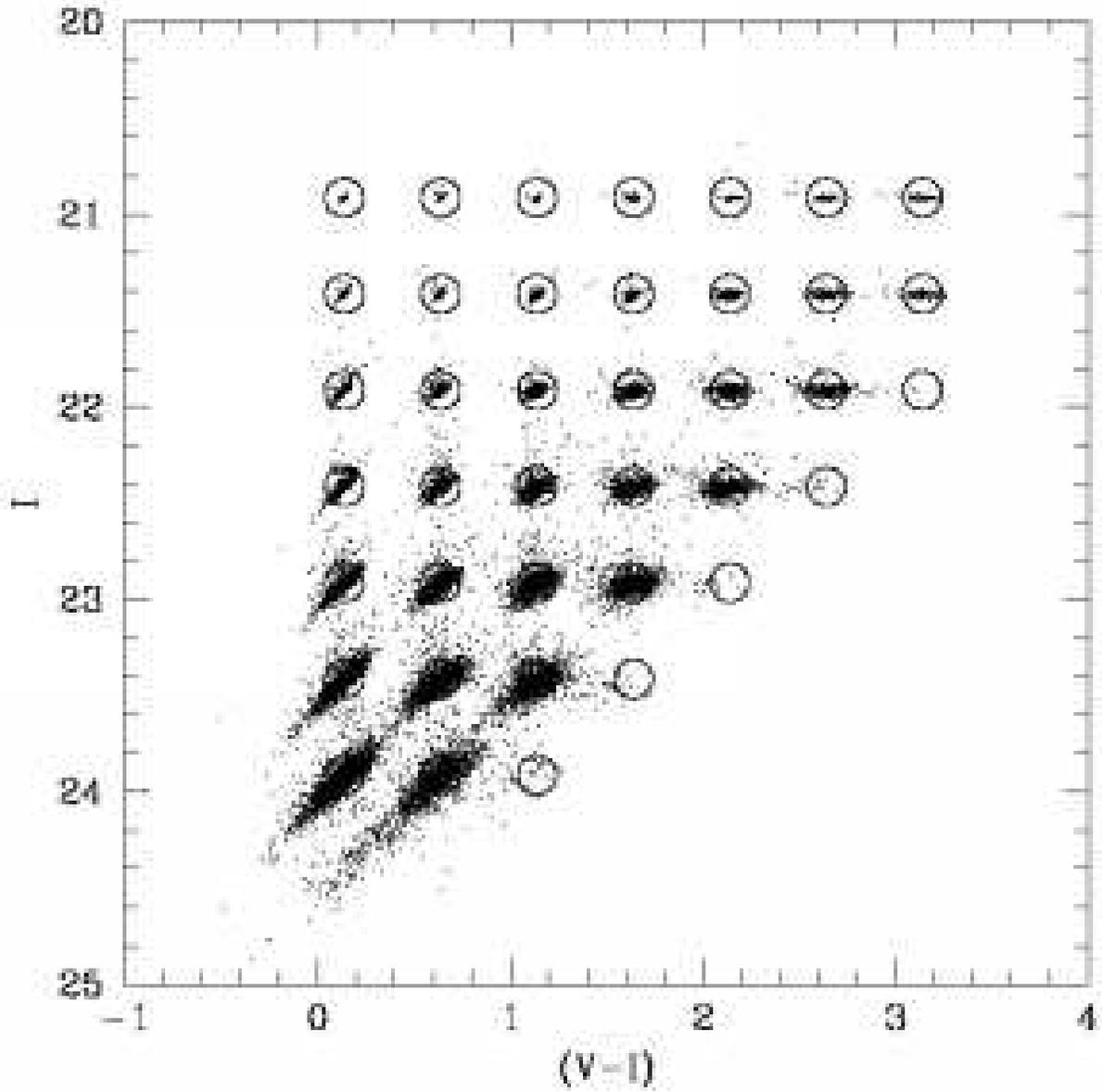}
\caption{An example of the color-magnitude diagram of stars recovered from the
artificial star tests. The circles are centered
on the locations of the added stars and the points represent the
recovered stars.\label{cmdcomp}}
\end{figure}

The integration of data from different chips in a CCD mosaic presents
some unique issues in characterization of the data. For example, the
quantum efficiencies of each chip differ, which results in varying
photometric uncertainties and incompleteness levels. To properly
quantify these effects, we have used the commonly employed method of
adding artificial stars of known magnitudes to the science frames and
re-reducing the images. In addition, to fully characterize the
incompleteness, the artificial star experiments were performed for a
range of $I$ magnitude and $V-I$ color. A total of 56,000 stars were
added to each chip in the mosaic over a series of 14 runs and over a
two dimensional grid on the color-magnitude diagram covering the range
$20.9 < I < 24.9$, $0.14 < (V-I) < 3.14$. The artificial star frames
were then reduced with the same procedures used for the original
science images. The color-magnitude diagram of recovered artificial
stars for one of the chips is shown in Figure~\ref{cmdcomp}.

\notetoeditor{Insert Figure~\ref{cmdcomp} here.}

The 50\% completeness level determined from the artificial star tests
is defined as the limiting magnitude, $m_{lim}$, and was derived by
fitting the data with the following interpolation function \citep{fle95}
\begin{equation} \label{interp}
f(m) = {\displaystyle \frac{\displaystyle 1}{\displaystyle 2}} \left[
1-{\displaystyle \frac{\alpha (m-m_{\rm
lim})}{\sqrt{(1+\alpha^{2}(m-m_{\rm lim})}}} \right]
\end{equation}
where $m$ represents the $V$ or $I$ magnitude and $\alpha$ is a shape
parameter that determines how sharply $f(m)$ decreases (typically 3.0
to 4.0 for this data). The limiting magnitudes $V_{lim}$ and $I_{lim}$
for each of the chips are summarized in Table~\ref{tbl-limmag}. The
photometric incompleteness for each individual star in
Figure~\ref{m33cmd} was obtained by linear interpolation from the
incompleteness $f(I,V-I)$ found for the grid points as determined by
the artificial star tests for each chip.

\section{The Luminosity Function and the Distance to M33}\label{sec-lf}

\begin{figure}
\plotone{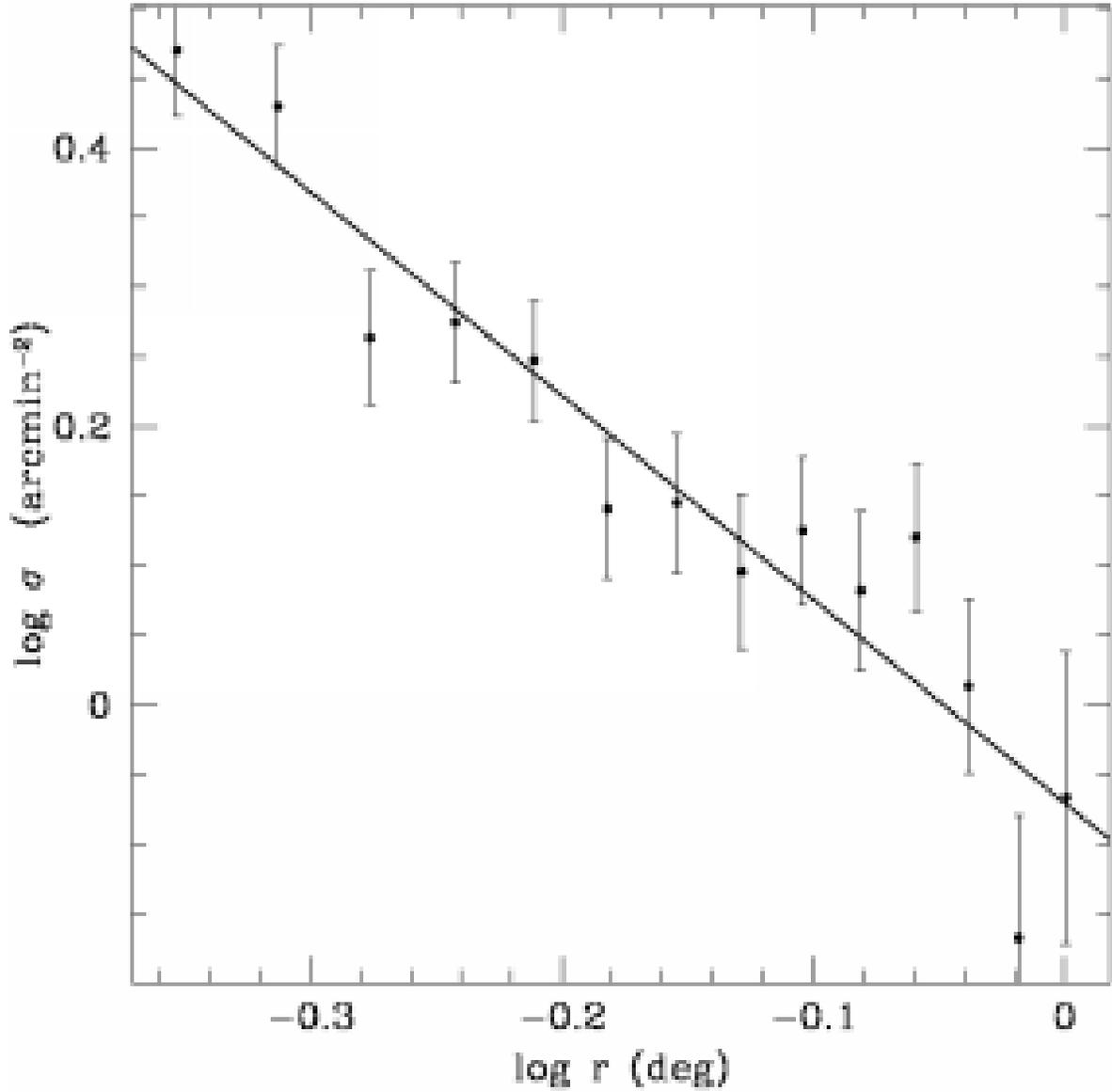}
\caption{Log-log plot of the stellar number density in the metallicity
distribution function with magnitudes in the range 20.5 $<$ I $<$ 22.5 as
a function of radius from the galactic center of M33. The solid line
is the least-squares best fit to the profile, with a slope of -1.46
$\pm$ 0.14.}\label{radial}
\end{figure}

The significant warp of M33's disk results in an abrupt radial cutoff
in the disk population. This cutoff was confirmed by examining star
counts as a function of radial distance from the center of M33. The
star counts drop precipitously across chips 00 and 06 (the two closest
to the nucleus of M33), which corresponds to the sharp terminus
exhibited by the HI disk \citep{rea78}. Figure~\ref{radial}
demonstrates that the star counts in the remaining 10 chips show a
clear dropoff as a function of radius in the M33 halo, assuring us
that we are indeed observing primarily a halo population and not a
foreground or background population.

\begin{figure}
\plotone{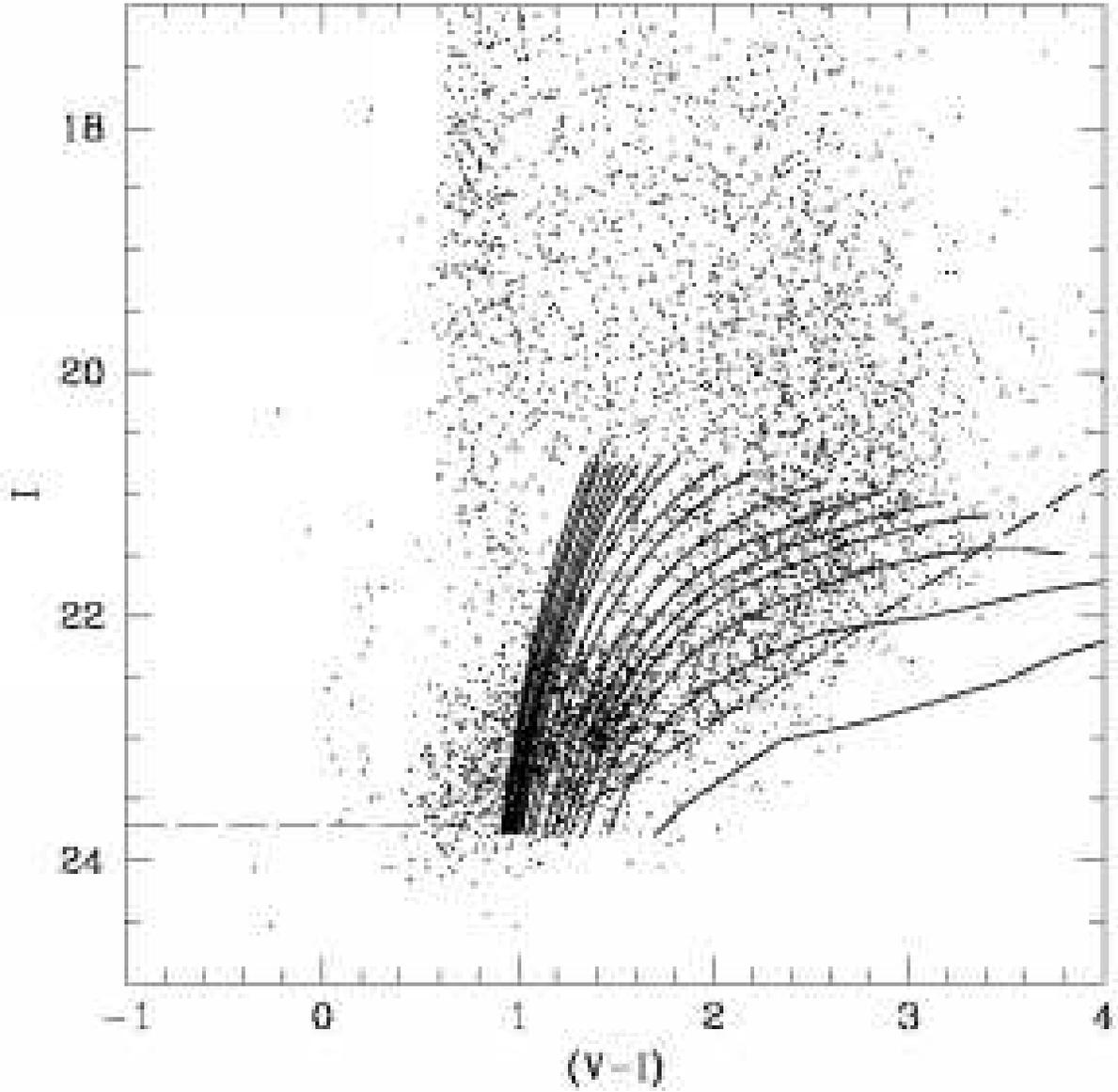}
\caption{Color magnitude diagram of adopted background field near M31,
from \citet{dur01}. The dashed line marks the 50\% completeness level and the solid lines show the theoretical models from \citet{van00}.\label{r1cmd}}
\end{figure}

To deal effectively with the problem of contamination due to faint,
unresolved background galaxies and foreground stars, we used a
background field from the study of M31 by \citet{dur01}. In examining
the stellar halo of M31, \citet{dur01} imaged a background field
(labeled $\mathscr R$1 in their paper) sufficiently distant from M31
to be free of M31 stars. The $\mathscr R$1 and M33 fields lie at
significantly different Galactic latitudes ($b=-21^\circ$ and
$b=-31^\circ$ respectively) and so the Galactic foreground
contamination differs between the two fields. The color
magnitude diagram for the $\mathscr R$1 field is shown in
Figure~\ref{r1cmd}; this color-magnitude diagram has also been cleaned
of nonstellar objects.

\notetoeditor{Insert Figure \ref{r1cmd} here}

The foreground reddening, $E(V-I)$, of the M33 field is also expected
to differ from that of the adopted background field. From the local HI
column densities from \citet{bur84}, the reddening of M33 is
$E(B-V)=0.04$. Adopting the relations $E(V-I)=1.35E(B-V)$ and
$A_I=1.95E(B-V)$ \citep{car89,bar00} yields a reddening of
$E(V-I)=0.054$ and extinction $A_I=0.078$ for M33. The reddening for
the $\mathscr R$1 field from the same source is $E(V-I)=0.08$, which
results in a differential reddening of $E(V-I)=0.026$ between the two
fields.

\begin{figure}
\plotone{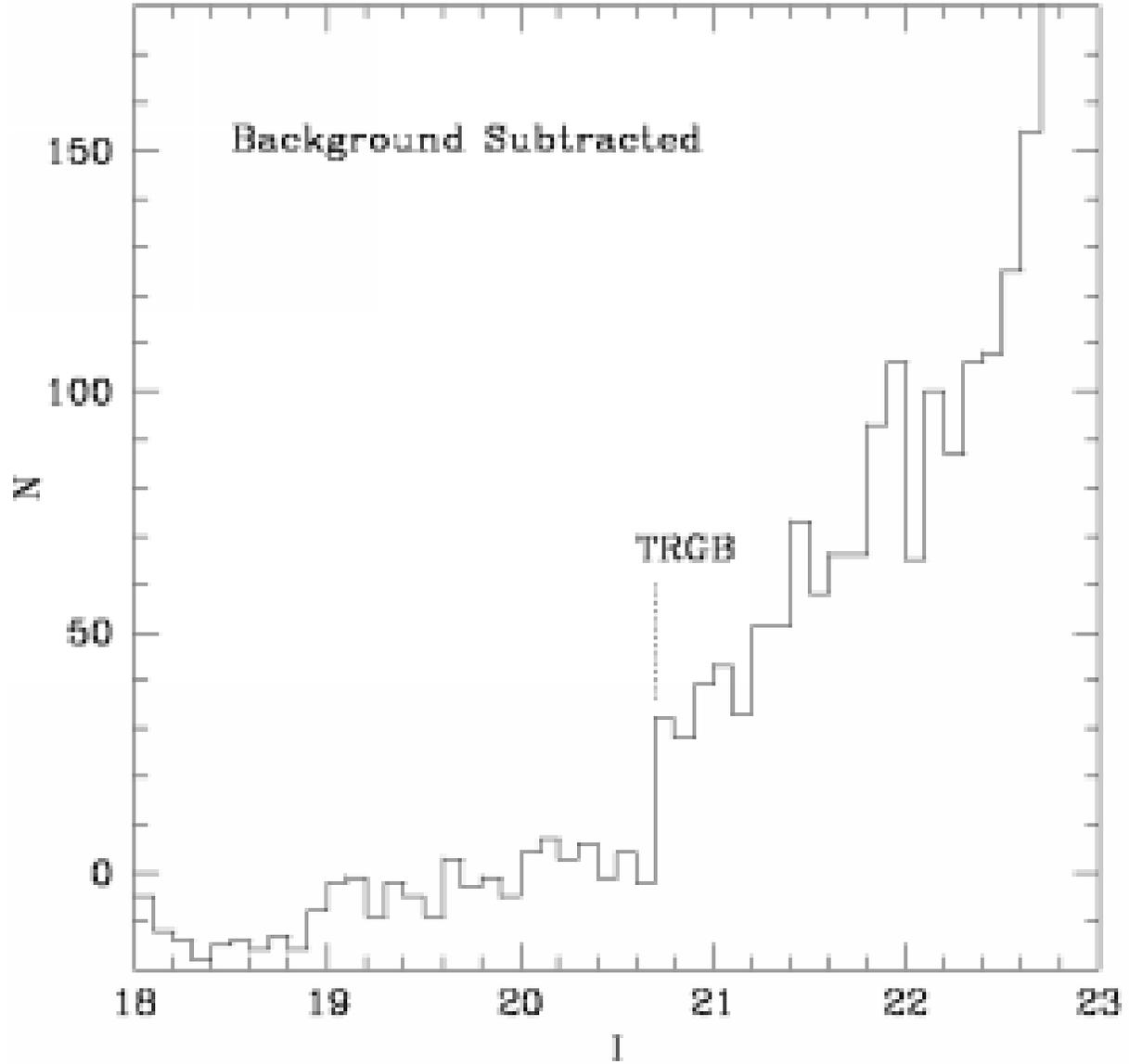}
\caption{The completeness corrected $I$-band luminosity function in the halo of M33. The location of the tip of the red giant branch is indicated; in the M33 halo it is seen at $I_{TRGB}=20.7$. Contamination by background galaxies has been removed as described in Section~\ref{sec-lf}.\label{lumfun}}
\end{figure}

The onset of core helium flash in low mass stars marks the bright tip
of the red giant branch. This sharp transition is evident as a sudden
upturn in the luminosity function of an old stellar population and so
can be used to determine accurate distances \citep{fro83}. Before
applying this method to the M33 halo field, it is useful to subtract
the $\mathscr R$1 luminosity function from that of the M33 halo. To
account for the differential reddening, $\Delta I= 0.042$ and
$E(V-I)=0.026$ were subtracted from the $\mathscr R$1 data. The
completeness-corrected luminosity function was then constructed by
applying the interpolated completeness factor, $f(I,V-I)$, to give
each star an effective count of $1/f$. Subtraction of the $\mathscr
R$1 field was dealt with by an iterative process to account for the
difference in Galactic foreground contamination (simply scaling the
two fields by their relative area overcorrects for foreground
contamination). Ideally, the luminosity function should average to
zero at the bright end, since this part of the luminosity function is
dominated by foreground Milky Way stars in both fields. We thus varied
the scaling factor over a range of 0.8 to 1.4 until the average number
counts were zero at the bright end of the luminosity function after
background subtraction. Stars over the magnitude range $19.0\leq I
\leq 20.5$ were used to derive this scale factor, which was then
applied to the $\mathscr R$1 luminosity function before subtraction
from the M33 halo luminosity function. The final completeness
corrected luminosity function for the M33 halo is shown in
Figure~\ref{lumfun}.

The magnitude of the tip of the red giant branch was found with an
edge-detection algorithm, which uses a numerical second derivative to
locate the abrupt change in slope associated with a sudden increase in
the number of stars at this magnitude. To test the sensitivity of this
method to the details of the background subtraction scaling, we used a
range of scale factors to perform the background subtraction and then
applied the edge detection algorithm to each case.

In all cases, the tip of the red giant branch was found repeatably at
the same $I$ magnitude, $I_{TRGB}=20.7\pm0.1$ mag. Adopting an
absolute magnitude turnoff of $M_{I,TRGB}=-4.1\pm0.1$ mag based on
Milky Way globular cluster data \citep{har98,har99} and using the
derived extinction ($A_I=0.078\pm0.029$ mag) yields a distance modulus
for M33 of $(m-M)_0=24.72\pm0.14$ mag. This distance modulus is in
good agreement with the recent HST $I$-band observations of Cepheids
in M33 by \citet{lee02}, which gives a distance modulus of
$(m-M)_0=24.52\pm0.14(random)\pm0.13(systematic)$. Other recent
results based on HST photometry of M33 disk stars in the $V$ and
$I$-bands give a distance modulus of
$(m-M)_0=24.81\pm0.04(random)^{+0.15}_{-0.11} (systematic)$
\citep{kim02} using the tip of the red giant branch.

\section{The Metallicity of the Halo of M33}\label{metal}

\subsection{The Stellar Metallicity Distribution Function}

Deriving the metallicity distribution function requires us to assume
that the M33 halo stars are old. The observed distribution of stars in
the color-magnitude diagram is a result of several factors, including
photometric scatter, a possible range of stellar ages, and a range of
stellar composition. The fiducial grid of red giant branch tracks used
in this work applies to old stars ($\tau>10$ Gyr); with this
assumption, only the metallicity of a model star determines its
location in the color-magnitude diagram. Evolutionary tracks from
\citet{van00} were used within an interpolation code \citep{har00} to
determine stellar metallicities.

The color-magnitude diagrams of the M33 halo and the $\mathscr R$1
fields are shown overlaid with the evolutionary tracks for 0.8
$M_{\sun}$ stars from \citet{van00}, which range from
$[$Fe$/$H$]=-2.31$ to $-0.04$. A single metal-rich isochrone
($[$Fe$/$H$]=+0.07$) was added to the model grid from \citet{ber94} to
allow the identification of any metal-rich halo stars. Following the
method of \citet{dur01} and \citet{har00}, the $(V-I)$ color of each
model was empirically shifted by -0.03 to match the observed red giant
branch distributions of Milky Way globular clusters over the full
metallicity range.

For the metallicity interpolation, the model grid was
first shifted by the distance modulus of M33 [$(m-M)_I=24.8$] and the
foreground reddening ($E(V-I)=0.054$). The metallicity of each individual
star was determined by bi-linear interpolation between evolutionary
tracks in the model grid. Each star was first transformed from the
$(M_I,(V-I))_{\circ}$ plane to the $(M_{bol},(V-I)_{\circ})$ plane
through the relation:
\begin{center}
$M_{bol} = M_{I} + (V-I)_{0} - BC_{I}$
\end{center}
where $BC_I$ is the bolometric correction for the star in the $I$-band
\citep{har00}. The bolometric corrections are provided with the
stellar models and these points are interpolated to generate the
bolometric correction for the star. The two evolutionary tracks
bracketing the star were identified and interpolation between these
tracks was performed within the $(M_{bol},(V-I)_{\circ})$ plane.

Metallicities were obtained for all of the stars in the
color-magnitude diagram over a magnitude range of 20.5 $<$ $I$ $<$
22.5 mag. Stars more than 0.02 mag bluer than the most metal-poor
track, as well as stars near the RGB tip (where interpolation is more
uncertain) were not used in the analysis.  At fainter $I$ magnitudes
($I > 22.5$), photometric uncertainties and field contamination
increase, and so the fainter end of the color-magnitude diagram was
not used for the metallicity distribution function analysis. At the
bright end above the red giant branch tip, the color-magnitude diagram
is dominated by contamination due to field stars that have survived
the background subtraction procedure.  Additionally, few M33 halo
stars are expected to be seen above the tip of the red giant branch;
the bright cutoff used in the derivation of the metallicity
distribution function reflects this.

\begin{figure}
\plotone{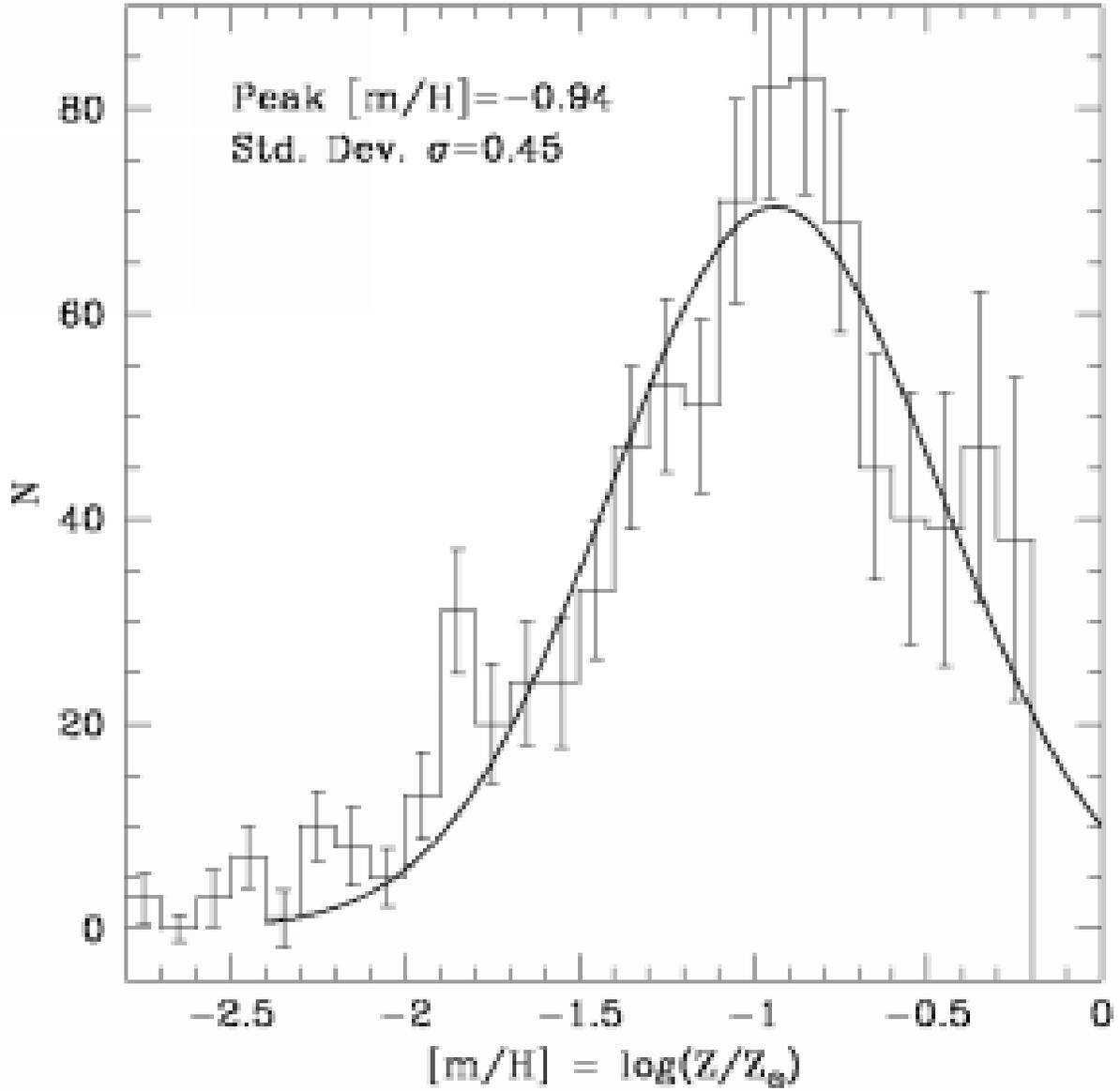}
\caption{The completeness corrected metallicity distribution function of the halo of M33. The best fit Gaussian is shown with a mean of $[$m/H$]=-0.94$. These counts have been corrected for background contamination. The error bars correspond to Poisson statistics.\label{mdf}}
\end{figure}

After metallicities were determined, the resulting star counts were
corrected for photometric incompleteness, background subtracted, and
then binned in 0.1 dex metallicity bins to give the distribution shown
in Figure~\ref{mdf}. A single component Gaussian was fit to this
metallicity distribution function, which identified a distinct,
metal-poor peak at $[m/$H]$=-0.94$ with a reduced chi squared of
$\chi^2=1.4$. The M33 halo population shows no indication of
multi-modality in its metallicity distribution function. A comparison
of the raw and background subtracted metallicity distribution
functions reveals that the peak at $[m/$H]$\approx-0.9$ is seen even
without background correction and thus is almost certainly due to the
halo of M33.

At the metal-rich end (at $[m/$H$]\geq-0.4$), the counts become more
uncertain and the background field may not adequately remove the
contamination from background galaxies. It is unlikely that we are
seeing a mix of halo and disk stars, because, as described in
Section~\ref{intro}, M33 has a significant disk warp which has the
effect of creating a rather sharp cutoff between the disk and halo.
Contamination may be present in the chips closest to M33; however,
this contamination is limited to chips 00 and 06, since the long axis
of the mosaic is oriented along the minor axis of M33. The combination
of these factors could well be responsible for the excess counts seen
in the most metal-rich bins. However, the remaining analysis will not be
significantly impacted by the uncertainties in these few metal-rich
bins.

At the metal-poor end ($[m/$H$]\leq-1.7$), there is expected to be a
population of asymptotic giant branch (AGB) stars which would make the
residual counts higher there. A close look at the lifetimes of the
stars along these two tracks over the range of metallicities used here
shows that, of the total sample, approximately $20\%\pm3\%$ stars will
be asymptotic giant branch stars [a similar result to that of
\citet{har99}]. Comparison of the AGB tracks and red giant branch
(RGB) tracks of \citet{gir00} over a range of metallicities and masses
reveals the AGB tracks to average $\approx0.1$ mag bluer in $(V-I)$
color than the RGB tracks. If we consider the AGB stars have the same
intrinsic metallicity peak as the RGB population, then the
metallicities of the aymptotic giant branch stars would be
interpolated to be $\approx0.15$ dex more metal poor than the RGB
stars. This effect would tend to broaden and bias the metallicity
distribution function, but only slightly.

\subsection{Comparison with the Globular Cluster System}

In comparison with the larger spirals of the Local Group, M33 has a
rather sparse globular cluster system, which has, however, been
studied much more extensively than the halo stars in M33. Christian \&
Schommer (1982, 1988) performed photometry on 130 candidate clusters
and identified 27 as possible old clusters ($> 10$ Gyr). Unlike the
globular cluster systems of M31 and the Milky Way, Christian \&
Schommer found M33 to have relatively massive clusters of all
ages. They concluded that cluster formation for globular clusters
younger than 10 Gyr was fairly continuous. Later work by \citet{sch91}
showed the ages of the globular clusters traced their kinematics, with
old globular clusters found in halo-like orbits and younger clusters
in disk-like orbits.

More recent studies of the globular cluster system of M33 using HST by
\citet{sar00} have measured metallicities for nine globular clusters
with ages $\geq4$ Gyr \citep{chr88} in the halo of M33. The current
sample of M33 globular clusters with measured metallicities is small
and so we can only compare the bulk properties of the cluster and halo
star populations. The mean metallicity of this sample of globular
clusters is $\langle$[Fe/H]$\rangle=-1.27\pm0.11$ dex, which is close
to the mean determined for the halo stars in this work,
$\langle$[Fe/H]$\rangle=-1.24\pm0.04$ dex (using [{\it m}/H] $\sim$
[Fe/H] + 0.3 from \citet{dur01}). There is also a significant spread
of metallicity in the halo globular clusters, similar to that seen in
the halo stars.

\subsection{Results from Chemical Evolution Models}

\begin{figure}
\plotone{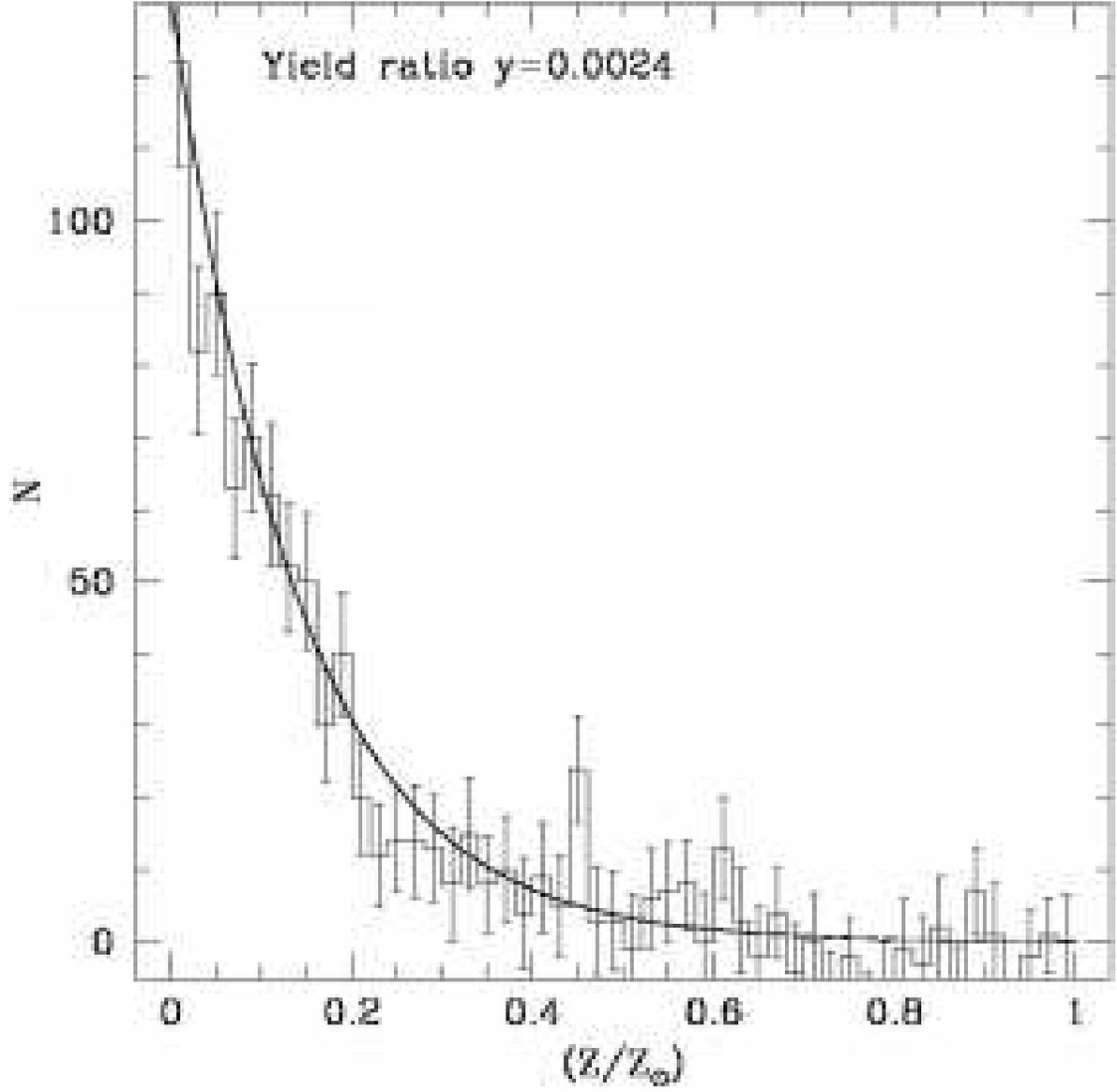}
\caption{Metallicity distribution function of M33 halo stars on a linear $(Z/Z_\sun)$ scale. The best-fit closed box model is the solid line, with a yield ratio of $y=0.0024$. The reduced chi squared of the fit is $\chi^2=1.1$.\label{logz}}
\end{figure}

We now compare the metallicity distribution function for the halo of
M33 with a simple chemical evolution model. In a ``closed-box''
chemical evolution model \citep{sea72,pag75}, the cumulative
distribution of stellar abundance, $Z$, is given by
\begin{equation}\label{eqn-diff}
{\displaystyle N(Z) \sim const~
\frac{1-e^{-(Z-Z_{\circ})/y}}{1-e^{-(Z_{N}-Z_{\circ})/y}}}
\end{equation}
where $N$ is the number of stars with a metal abundance less than $Z$,
$Z_\circ$ is the initial metal abundance of the gas, $Z_N$ is the
current metal abundance of the gas, and $y$ is the yield ratio [the
mass of gas in metals expelled by massive stars divided by the mass of
dead stellar remnants \citep{pag75}]. ``Leaky-box'' models use an
effective yield, $y_{eff}$, given in terms of the yield ratio and a
parameter $c$ that describes the fraction of mass lost from the box,
that is, $y_{eff}=y/(1+c)$. The mass loss is thought to be driven by
supernovae and winds from massive stars, which shock the resident gas,
expelling it from the idealized closed box model \citep{har76}. The
functional form of the differential abundance distribution remains the
same with $y$ replaced by $y_{eff}$ in equation~\ref{eqn-diff}.  When
fitting the leaky-box model to the M33 data, we use the differential
distribution $dN/dZ$
\begin{equation}
{\displaystyle \frac{dN}{dZ} \sim const~
\frac{e^{-(Z-Z_{\circ})/y_{eff}}}{y_{eff}(1-e^{-(Z_{N}-Z_{\circ})/y_{eff}})}}
\end{equation}
Since the expression is in terms of $Z$, it is more natural to express
the metallicity distribution function as a linear function of
$(Z/Z_\sun)$, as shown in Figure~\ref{logz}.  \notetoeditor{Insert
Figure~\ref{fig-logz} here.}  We kept the initial and current metal
abundance fixed during the model fitting ($Z_\circ=0$ and
$Z_N=0.0172$, respectively). The best-fit value of the effective
yield, $y_{eff}$, was found to be $y_{eff}=0.0024\pm0.0002$ and the
resulting fit is represented by the solid line in
Figure~\ref{logz}. Bin-to-bin deviations exist, but given the size of
the error bars, these are not significant. The leaky-box model proves
to give an excellent fit to the data.

A similar analysis using a ``leaky box'' model has been performed for
M31 \citep{dur01} and the Milky Way \citep{rya91}. The value of the
effective yield for the M33 halo clearly lies between the values for
M31 and the Milky Way. \citet{rya91} find best fit values of $y_{eff}$
to be $\sim0.0009$ for the Milky Way's halo, while \citet{dur01} find
a best fit value of $y_{eff}=0.005$ for M31. If the Milky Way and M31
had similar intrinsic yields, their $c$ values would then differ
considerably, which would imply that the Milky Way expelled a much
larger fraction of the initial gas in its halo during its assembly. By
the same token, M33 was not as efficient as the Milky Way in expelling
its gas, but was about twice as efficient as M31. Thus, the enrichment
of the halo of M33 was able to proceed to a greater degree than in the
Milky Way. Given the relatively shallow gravitational potential well
of M33, it is surprising that M33 was better at retaining its gas than
was the Milky Way.

Early studies of the halo of our Galaxy by \citet{egg62} argued for a
single, radial collapse of a proto-Galactic gas cloud. This process
would leave a distinct metallicity gradient, since the formation of
stars during the collapse would proceed from increasingly metal-rich
gas. \citet{sch93} found some evidence for a radial metallicity
gradient among old globular clusters. However, \citet{sar00} found no
support for this conclusion; in both cases, the sample sizes are quite
small. If M33 does, in fact, lack a metallicity gradient, this would
support the idea that M33 was formed by a build-up of smaller,
spheroidal systems rather than a single, radial
collapse. Unfortunately, the size of our stellar sample does not
permit a statistically reliable constraint on the presence or absence
of a metallicity gradient in the halo of M33.

\section{Summary}

We have presented the results of $(I,V-I)$ photometry of the halo
stars of M33 in a field located 10 kpc from the galaxy center along
its southeast minor axis. The stellar halo population was isolated
through the use of image classification and a background field
provided by \citet{dur01}. With an assumed reddening of
$E(V-I)=0.054\pm0.020$, the distance modulus based on the tip of the
red giant branch was found to be $(m-M)_\circ=24.72\pm0.14$ mag.

Through the use of a grid of red giant branch stellar models,
metallicity values for individual stars were interpolated allowing the
derivation of the halo metallicity distribution function. The
metallicity distribution function shows a distinct peak at
$[m/$H$]=-0.94\pm0.042$ (or $\langle$[Fe/H]$\rangle=-1.24\pm0.04$ dex
(using [{\it m}/H] $\sim$ [Fe/H] + 0.3 from \citet{dur01}) and agrees
well with the mean metallicity measured for nine M33 globular clusters
of $\langle$[Fe/H]$\rangle=-1.27\pm0.11$ by \citet{sar00}.

A single component leaky box chemical evolution model provides a good
fit to the metallicity distribution, resulting in a yield of
$y_{eff}=0.0024$. Comparison of the yield of M33 with those of M31 and
the Milky Way ($y_{eff}=0.0009$ and $0.005$, respectively) places the
M33 yield between these values. It is somewhat surprising that M33 has
a higher effective yield compared to the Milky Way, since it has a
significantly smaller potential well to confine its gas. However, if
the halos of these galaxies were built up from the accretion of
smaller satellites, an alternate interpretation would be that the
subsystems were smallest for the Milky Way, larger for M33, and still
larger for M31 (from the standard mass-metallicity relation for dwarf
galaxies).

\acknowledgments
We would like to thank Pat Durrell for making available the M31
background data and for several fruitful discussions concerning the
reduction of CFHT12K data. This research was supported by NSERC
research grants to C. D. W. and W. E. H., and R. S. B. gratefully
acknowledges financial support from C. D. W. We also acknowledge the
use of archival data made available by the Canadian Astronomy Data
Centre, which is operated by the Herzberg Institute of Astrophysics,
National Research Council of Canada.

\clearpage

\clearpage 

\begin{deluxetable}{cccc}
\tabletypesize{\scriptsize}
\tablecaption{Transformation Coefficients \label{tbl-values}}
\tablewidth{0pt}
\tablehead{
\colhead{} & \colhead{$a_n$} & \colhead{$b_n$} & \colhead{$c_n$} }
\startdata
V & 1.319 $\pm$ 0.003 & 0.016 $\pm$ 0.003 & -0.152 \\
I & 1.012 $\pm$ 0.005 & -0.016 $\pm$ 0.005 & -0.061 \\
\enddata
\end{deluxetable}

\clearpage

\clearpage

\clearpage

\clearpage

\clearpage

\begin{deluxetable}{cccccc}
\tabletypesize{\scriptsize}
\tablecaption{Limiting Magnitudes\label{tbl-limmag}}
\tablewidth{0pt}
\tablehead{
\colhead{CCD No.} & \colhead{$V_{lim}$} & \colhead{$I_{lim}$} & \colhead{CCD No.} & \colhead{$V_{lim}$} & \colhead{$I_{lim}$} }
\startdata
01 & 24.67 & 23.59 & 07 & 24.72 & 23.92 \\
02 & 24.76 & 24.06 & 08 & 24.71 & 23.92 \\
03 & 25.03 & 24.25 & 09 & 24.85 & 23.96 \\
04 & 24.62 & 23.95 & 10 & 24.72 & 23.85 \\
05 & 24.84 & 24.01 & 11 & 24.73 & 23.95 \\
\enddata
\end{deluxetable}

\clearpage

\clearpage

\clearpage

\clearpage

\begin{thebibliography}{}

\bibitem[Barmby et al.(2000)]{bar00} Barmby, P., Huchra, J. P.,
Brodie, J. P., Forbes, D. A., Schroder, L. L., \& Grillmair, C. J.,
2000, \aj, 119, 727

\bibitem[Bertelli et al.(1994)]{ber94} Bertelli, G., Bressan, A.,
Chiosi, C., Fagotto, F., and Nasi, E., 1994, A\&AS, 106, 275

\bibitem[Burstein \& Heiles(1984)]{bur84} Burstein, D., \& Heiles, C.,
1984, \apjs, 54 33

\bibitem[Cardelli et al.(1989)]{car89} Cardelli, J. A., Clayton,
G. C., \& Mathis, J. S., 1989, \apj, 345, 245

\bibitem[Christian \& Schommer(1982)]{chr82} Christian, C. A.,
Schommer, R. A., 1982, \apjs, 49 405

\bibitem[Christian \& Schommer(1988)]{chr88} Christian, C. A.,
Schommer, R. A., 1988, \aj, 95, 704

\bibitem[Deul \& van der Hulst(1987)]{deu87} Deul, E. R., van der
Hulst, J. M., 1987, A\&AS, 67, 509

\bibitem[Durrell et al.(2001)]{dur01} Durrell, P. R., Harris, W. E.,
and Pritchet, C. J., 2001, \aj, 121, 2557

\bibitem[Eggen et al.(1962)]{egg62} Eggen, O., Lynden-Bell, D., \&
Sandage, A. R., 1962, \aj, 136, 748

\bibitem[Fleming et al.(1995)]{fle95} Fleming, D. E. B., Harris,
W. E., Pritchet, C. J., \& Hanes, D. A., 1995, \aj, 109, 1044

\bibitem[Frogel et al.(1983)]{fro83} Frogel, J. A., Cohen, J. G.,
Persson, \& S. E., 1983, \apj, 275, 773

\bibitem[Girardi et al.(2000)]{gir00} Girardi, L., Bressan, A.,
Bertelli, G., \& Chiosi, C., 2000, A\&AS, 141, 371

\bibitem[Harris et al.(1981)]{har81} Harris, W. E., Fitzgerald, M. P.,
\& Reed, B. C. 1981, PASP, 93, 507

\bibitem[Harris \& Harris(2000)]{har00} Harris, G. L. H., Harris,
W. E., 2000, \aj, 120, 2423

\bibitem[Harris et al.(1999)]{har99} Harris, G. L. H., Harris, W. E.,
and Poole, G. B., 1999, \aj, 117, 855

\bibitem[Harris et al.(1998)]{har98} Harris, W. E., Durrell, P. R.,
Pierce, M. J., and Secker, J., 1998, Nature, 395, 45

\bibitem[Harris(2001)]{har01} Harris, W. E., 2001, in Star Clusters,
Saas-Fee Advanced Course 28 (New York: Springer), ed. L.Labhardt \&
B.Binggeli

\bibitem[Hartwick(1976)]{har76} Hartwick, F. D. A., 1976, ApJ, 209, 418

\bibitem[Kim et al.(2002)]{kim02} Kim, M., Kim, E., Lee, M. G.,
Sarajedini, A., and Geisler, D., 2002, \aj, 123, 244

\bibitem[Kron(1980)]{kro80} Kron, R. G., 1980, \apjs, 43, 205

\bibitem[Landolt(1992)]{lan92} Landolt, A. U., 1992, \aj, 104, 340

\bibitem[Lee et al.(2002)]{lee02} Lee, M. G., Kim, M., Sarajedini, A.,
Geisler, D., \& Gieren, W., 2002, \apj, 565, 959

\bibitem[Mould \& Kristian(1986)]{mou86} Mould, J., \& Kristian, J.,
1986, \apj, 305, 591

\bibitem[Pagel \& Patchett(1975)]{pag75} Pagel, B. E. J., \& Patchett,
B. E., 1975, \mnras, 172, 13

\bibitem[Reakes \& Newton(1978)]{rea78} Reakes, M. L., Newton, K.,
1978, \mnras, 185, 277

\bibitem[Rogstad et al.(1976)]{rog76} Rogstad, D. H., Wright,
M. C. H., and Lockhart, I. A., 1976, \apj, 204, 703

\bibitem[Ryan \& Norris(1991)]{rya91} Ryan, S. G., \& Norris, J. E.,
1991, \aj, 101, 1865

\bibitem[Sarajedini et al.(2000)]{sar00} Sarajedini, A., Geisler, D.,
Schommer, R., Harding, P., 2000, \aj, 120, 2437

\bibitem[Schommer(1993)]{sch93} Schommer, R. A., 1993, ASP
Conf. Ser. 48, The Golbular Cluster-Galaxy Connection, ed. G. H. Smith
\& J. P. Brodie (San Francisco: ASP), 458

\bibitem[Schommer et al.(1991)]{sch91} Schommer, R. A., Christian,
C. A., Caldwell, N., Bothun, G. D., \& Huchra, J., 1991, \aj, 101, 873

\bibitem[Searle \& Sargent(1972)]{sea72} Searle, L., \& Sargent,
W. L., 1972, \apj, 173, 25

\bibitem[Stetson(1987)]{ste87} Stetson, P. B., 1987, \pasp, 99, 191

\bibitem[Stetson(1992)]{ste92} Stetson, P. B., 1992, in ASP
Conf. Ser. 25, Astronomical Data Analysis Software and Systems I,
ed. D. M. Worral, C. Biemesderfer, \& J. Barnes (San Francisco: ASP),
297

\bibitem[Stetson, Davis \& Crabtree(1990)]{ste90} Stetson, P. B., Davis,
L. E. \& Crabtree, D. B., 1990, in CCDs in Astronomy, ASP Conf. Ser. 8,
Jacoby G.H., (ASP, San Francisco), p.289

\bibitem[Stetson(1989)]{ste89} Stetson, P. B., 1989, in Some Factors
Affecting the Accuracy of Stellar Photometry with CCDs (And Some Ways
of Dealing with them), in Highlights of Astronomy, 8, 635,
ed. D. McNally

\bibitem[VandenBerg et al.(2000)]{van00} VandenBerg, D. A., Swenson,
F. J., Rogers, F. J., Iglesias, C. A., and Alexander, D. R., 2000,
\apj, 532, 430

\end{thebibliography}
\end{document}